\documentclass[aps,twocolumn,pra,english,amssymb,amsmath,showpacs,superscriptaddress]{revtex4-1}

\usepackage{ucs}
\usepackage[english]{babel}
\usepackage{amsmath}
\usepackage{amssymb,amsfonts}
\usepackage{color}

\usepackage{graphicx}
\usepackage{dcolumn}
\usepackage{bm}

\usepackage{bbold}

\usepackage{subfigure}

\usepackage{tikz}
\usetikzlibrary{shapes.misc}

\usepackage{mathrsfs}

\usepackage{scrpage2}
\usepackage{hyperref}
\newcommand{\bd}{b^{\dagger}}
\newcommand{\bv}{b}
\newcommand{\dBd}{\Delta B^{\dagger}}
\newcommand{\dBv}{\Delta B}

\newcommand{\omegaL}{\ensuremath{{\omega_{\mathrm{L}}}}}

\begin{document}

\title{Quantum Pyragas control: Selective-control of individual photon probabilities}
\author{Leon Droenner}%
\affiliation{Technische Universit\"at Berlin, Institut f\"ur Theoretische Physik, Hardenbergstra{\ss}e 36, 10623 Berlin, Germany}
\author{Nicolas L. Naumann}
\affiliation{Technische Universit\"at Berlin, Institut f\"ur Theoretische Physik, Hardenbergstra{\ss}e 36, 10623 Berlin, Germany}
\author{Eckehard Sch\"oll}
\affiliation{Technische Universit\"at Berlin, Institut f\"ur Theoretische Physik, Hardenbergstra{\ss}e 36, 10623 Berlin, Germany}
\author{Andreas Knorr}
\affiliation{Technische Universit\"at Berlin, Institut f\"ur Theoretische Physik, Hardenbergstra{\ss}e 36, 10623 Berlin, Germany}
\author{Alexander Carmele}
\email{alex@itp.tu-berlin.de}
\affiliation{Technische Universit\"at Berlin, Institut f\"ur Theoretische Physik, Hardenbergstra{\ss}e 36, 10623 Berlin, Germany}

\begin{abstract}
Pyragas control allows to stabilize unstable states in applied
nonlinear science.
We propose to apply a quantum version of the Pyragas protocol to control individual photon-probabilities in an otherwise only globally accessible photon-probability distribution of a quantum light emitter. 
The versatility of quantum Pyragas control is demonstrated for the case of a two-level emitter in a pulsed laser-driven half cavity. 
We show that one- and two-photon events respond in a qualitatively different way to the half-cavity induced feedback signal.
One-photon events are either enhanced or suppressed,
depending on the choice of parameters. 
In contrast, two-photon events undergo exclusively an enhancement up to $50\%$ for the chosen pulse areas. 
We hereby propose an implementation of quantum Pyragas control via a time-delayed feedback setup. 
\end{abstract}

\maketitle
\section{Introduction}
Since its introduction, the delayed feedback control method \cite{bechhoefer2005feedback,scholl2008handbook}
is still one of the most active fields in applied nonlinear science \cite{isidori2013nonlinear,boccaletti2000control,khalil2015nonlinear,scholl2016control}. 
Pyragas control is a specific form of such a closed-loop feedback control protocol which allows to force non-invasively a system into a desired target state and vanishes as soon as this state is attained \cite{pyragas1992continuous}.
Being reference signal-free the controlled system can be treated as a black box as 
no exact knowledge of either the form of the periodic orbit or the system of equations is
needed. 
A standard (classical) Pyragas control takes the form:
\begin{align}
\dot x(t) =& f(x(t),t) - K \left[ x(t) - x(t-\tau) \right].
\end{align}
Hence, whenever the delay $ \tau $ is an integer multiple of the period of the target solution 
$ x(t) = x(t+\tau) $ of the uncontrolled nonlinear system $\dot x=f(x)$, the solution persists and the control force $ K $ vanishes
on the target orbit.
Experimental successful implementations of the Pyragas method include, e.g. control of unstable orbits in CO$_2$ laser with modulated losses \cite{bielawski1994controlling} and has a wide range of applications in semiconductor laser systems \cite{schikora2006all,fiedler2008delay,schikora2011odd,erneux2010laser,oliver2015consistency}. %
In electronic systems, time-delayed feedback is applied to enforce
autosynchronization in diode resonators \cite{gauthier1994stabilizing,unkelbach2003time}, in chemical systems to control chaos in Belousov-Zhabotinsky reactions \cite{schneider1993continuous}, and addresses birhythmicity in physical, biological systems in a noninvasive way \cite{biswas2016control,banerjee2017time}.
In the physics of plasma, Pyragas method has been employed
to control current-driven ion acoustic instabilities \cite{fukuyama2002dynamical} and unstable low-frequency electrostatic waves arising from strong modulations of ion and electron densities \cite{gravier2000dynamical}.
\begin{figure}[b!]
\centering
\includegraphics[width=\columnwidth]{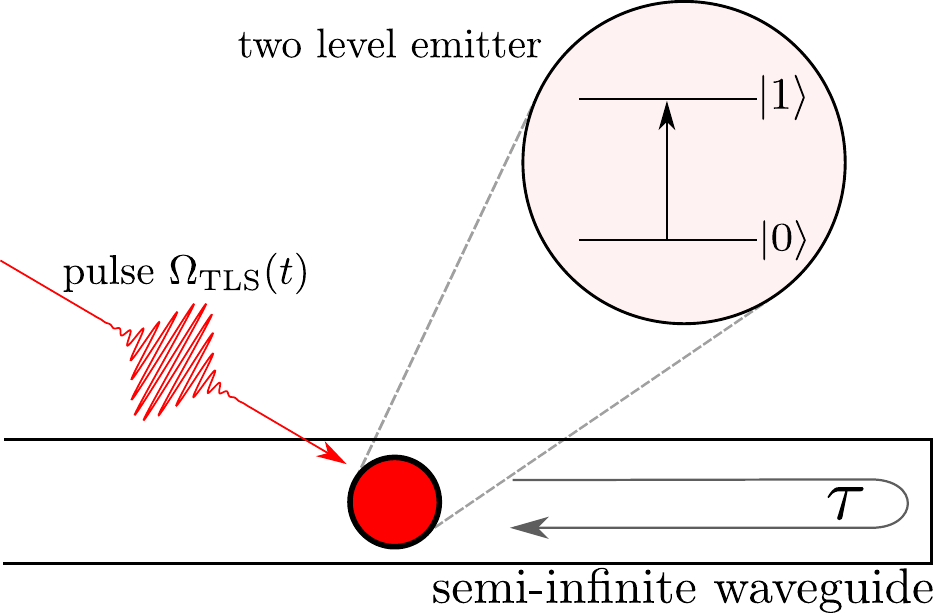}
\caption{Illustration of the system under pulsed excitation where a single TLS is placed inside a waveguide. A photon propagating to the right side is reflected and may excite the TLS again with delay $\tau$. }
\label{fig:setup}
\end{figure}
Despite the successes in semiclassical and classical nonlinear
systems, feedback control in the quantum regime has been mostly
investigated only in open-loop control, i.e. measurement-based protocols
\cite{quantumnoise,koch2016controlling} with successes e.g. in Fock-state preparation in microwave cavity-QED platforms \cite{sayrin2011real}
or persistent control of superconducting qubits \cite{campagne2013persistent}.
Lately, considerable interest shifted to closed-loop feedback control \cite{wiseman1,zhang2017quantum}
and based on various theoretical models \cite{carmele2013,PhysRevLett.115.060402,PhysRevLett.116.093601,Kabuss:16} predictions include stabilization of Rabi oscillations in the presence of a structured reservoir \cite{carmele2013}, control of unstable branches of bistable optomechanical systems \cite{wang2017enhancing,PaperOptoControl2014,rossi2017enhancing}, synchronization of network nodes 
\cite{PhysRevA.91.052321,PhysRevLett.116.093601,gough2014feedback,cook2018input}, enhancement of polarization-entanglement in a biexciton cascade \cite{hein14}, antibunching in multi-photon cavity-QED \cite{PhysRevA.95.063840}, and squeezing 
in parametric oscillators \cite{PhysRevA.94.023806, PhysRevA.94.023809}.
In addition to these examples, we propose here
a completely novel type of quantum control, allowing to stabilize 
a single photon-probability in the photon-probability distribution of a quantum light emitter without changing neighboring probabilities. 
The system we propose is based on all-optical quantum feedback of a two-level system (TLS) which is driven by an external pulsed laser field, cf. Fig. \ref{fig:setup}, and in which the pulse-area controls the emission characteristics. 
To trigger single-photon emission, the Gaussian pulse inverts 
the TLS ($\pi$-pulse), and a single photon is emitted subsequently due to radiative relaxation.
However, if a $ 2\pi-$pulse is applied, the TLS favors 
a two-photon emission as has been demonstrated lately 
theoretically and experimentally \cite{1367-2630-16-5-055018, nphys4052, fischer2017pulsed}.
The goal of this study is to demonstrate that Pyragas quantum feedback control is able to selectively suppress,
enhance and mediate between one- and two-photon emission events.
Selective control here means that non-Markovian feedback allows
to enhance a single photon-probability without affecting other
photon-probabilities. 
Typical measurement-based quantum control is modeled via a 
Lindblad-type jump operator acting on the full system density matrix $ \mathcal{D}[J]\rho=2J\rho J^\dagger-\lbrace J^\dagger J,\rho\rbrace $.
Given a Markovian system-environment coupling, the dynamics of a single photon probability $p(n)=\left\langle n \right| \rho \left| n \right\rangle $ reads with photon annihiliation and creation operator $b$ and $ b^{\dagger} $, respectively, and decay constant $ \kappa $: 
\begin{align*}
\dot p(n)/\kappa
=& 
\left\langle n \right|  \mathcal{D}[b]\rho \left| n \right\rangle 
= 
-2np(n)+2(n+1)p(n+1),
\end{align*}
in which necessarily the photon probabilities couple to each other
due to the quantum jump part of the Lindblad-operator.
In contrast, as we demonstrate now, in a non-Markovian quantum 
control setup, we are able to address just a single photon-probability and thus enhance $ p(n) $ without influencing $ p(n+1) $ or $ p(n-1) $.
We hereby expand the potential range of Pyragas control based on its quantum regime analogue:
\begin{align}
\dot x(t) =& 
\frac{i}{\hbar}
\left[H_s,x(t)\right]
-K\left(
x(t)-e^{i\phi}
x(t-\tau)
\right)
+N(t)
\label{eq:qPyragas},
\end{align}
for the Heisenberg operator $ x(t) $ and $ t>\tau $ with 
system dynamics induced by the system's Hamiltonian  $ H_s $
and the Pyragas control contribution \cite{carmele2013,Kabuss:16}.
Note that the quantum version of Pyragas control includes 
inevitably a control phase parameter $ \phi $ and 
due to the control environment a noise operator $ N(t) $
to ascertain the conservation of the canonical commutation relation of $ [x,p]=i\hbar$ \cite{quantumnoise}.
As a physical implementation we have in mind a light reflecting element (external mirror, integrated semi-infinite waveguide) 
which feeds back the photons emitted by the system back into the 
system after a roundtrip of $ \tau $, cf. Fig. \ref{fig:setup}.
This is a quantum version of the Lang-Kobayashi setup 
\cite{LAN80b} and has already been realized in the 
quantum regime for cold atoms and semiconductor lasers
\cite{PhysRevLett.96.043003,nc6}.
Due to the mirror-induced boundary condition, the dynamics is essentially non-Markovian and due to the driving laser-field standard quantum optical methods fail to model the system. 
Here, we model the feedback with the quantum stochastic Schr\"odinger equation \cite{quantumnoise,PhysRevLett.95.110503, PhysRevLett.116.093601}, where a matrix product state (MPS) representation allows to treat only the most relevant part of the Hilbert space corresponding to a numerically exact
treatment \cite{manzoni2017simulating,strathearn2018efficient}.
\section{Quantum Pyragas model}
We consider a single TLS with transition energy $ \hbar\omega_0 $ inside a semi-infinite waveguide \cite{PhysRevLett.98.083603, PhysRevA.91.053845}, cp. Fig. \ref{fig:setup}.
A spontaneous decay of the electronic excited state induced by the lowering operator $\sigma_-|1\rangle=|0\rangle$ emits a photon into the waveguide. 
The waveguide is closed at the right side, for instance by a reflecting cavity, acting as a mirror.
We model the interaction between the waveguide and the TLS with the following, quantum feedback-inducing Hamiltonian:
\begin{align}
H_{\mathrm{fb}}
=
\hbar g_0 
\int_{\mathcal{B}}d\omega \left[\sin(\omega L/c_0) \bd(\omega) \sigma_-  + \text{h.a.}\right],
\label{eq:Hfb}
\end{align}
with $b^{(\dagger)}(\omega)$ being the annihilation (creation) operator for a waveguide photon of frequency $\omega$
and raising/lowering operator for atomic excitation $ \sigma_{\pm}$. 
The TLS-reservoir interaction, is described by $G_{\mathrm{fb}}(\omega)=g_0 \sin(\omega L/c_0)$ \cite{PhysRevA.93.053807, doi:10.1080/09500340.2017.1363919} with $c_0$ as the speed of light in the waveguide. 
The coupling $G_{\mathrm{fb}}(\omega)$ includes the reflecting mirror at distance $L$ from the TLS with time-delay $\tau=2L/c_0$ before an emitted photon again interacts with the TLS. 
This interaction Hamiltonian $H_{\mathrm{fb}}$ gives rise to the quantum Pyragas equation for a given waveguide photon or system operator in the Heisenberg picture, cf. Eq.~\eqref{eq:qPyragas}.
\begin{align}
\notag
\dot \sigma_-(t) =
&
\frac{i}{\hbar}[H_s,\sigma_-] 
-
\Gamma 
\sigma_-(t)
+ N(t)\\
\notag
&
+\Gamma
e^{i\omega_{0}\tau}
\sigma_-(t-\tau)
\sigma_+(t)\sigma_-(t)
\theta(t-\tau)\\
&
-
\Gamma 
e^{i\omega_{0}\tau}
\sigma_-(t-\tau)
\sigma_-(t)\sigma_+(t)
\theta(t-\tau)
\end{align}
where we have set $ \Gamma=\pi g_0^2/2$ as the radiative 
decay constant and $N(t) =i\int_{\mathcal{B}}d\omega 
G_{\mathrm{fb}}(\omega) \bd_0(\omega) \exp[-i\omega t]$
denotes the noise contribution which conserves the commutation 
relation and $\bd_0(\omega)$ as the annihilation operator
of a waveguide photon at $ t=0$.
If $ t>\tau$ and $ \sigma_-(t)=\sigma_-(t-\tau) $, the equation of motion reduces to 
\begin{align}
\notag
&\dot \sigma_-(t)|_\text{periodic} 
=
\frac{i}{\hbar}[H_s,\sigma_-]
-\Gamma 
\left[
1
+
e^{i\omega_{0}\tau}
\right] 
\sigma_-(t)
+N(t)
\end{align}
and for specific phases $ \phi=\omega_0\tau=n\pi$ for $ n$ integer and negligible noise contributions, we recover the pure system dynamics governed by $H_s$ as in the classical case.
Due to the phase and quantum noise contributions, the quantum version of the Pyragas method offers new degrees of freedom beyond control of periodic orbits. 
In the following, we show that exactly this phase via a given delay time allows to address selectively a single photon-probability in a photon-probability distribution $ p(n) $.
\\
\section{Quantum feedback in the matrix product state picture}
We consider that the system dynamics is externally controlled
via an external coherent pulse, resonant with the TLS-frequency $\omega_{0}$. 
As a control parameter, we choose the pulse area.
The external laser is modeled as a Gaussian-pulse with frequency $\omega_L$ and amplitude, 
$
\Omega(t)=A \ e^{-t^2/\nu^2}/\sqrt{\nu^2\pi},
$
giving rise to the pulse area $A$ in terms of the temporal width of the pulse  $\nu$.
We choose the pulse to be short in comparison with the inverse decay rate of the electronic excited state $\Gamma$ in the same manner as in Ref. \cite{nphys4052}. For a longer pulse duration, probabilities of higher photon numbers would become more relevant which is beyond the scope of this present study.
The total Hamiltonian reads:
$H_{\mathrm{tot}}=H_0+H_{\mathrm{s}}(t)+H_{\mathrm{fb}}$ with
\begin{align}
H_0 &= 
\hbar \omega_{0} \sigma_+ \sigma_-
+
\int_{\mathcal{B}}d\omega\hbar\omega\bd(\omega)\bv(\omega) \\
H_{\mathrm{s}}(t)&=
\hbar \Omega(t) \left(  \sigma_+ e^{-i\omegaL t} + \sigma_- e^{i\omegaL t} \right) \label{eq:HTLS}
\end{align}
being the Hamiltonian of the pumped TLS in energy conserving rotating wave and dipole approximation. 
Furthermore, we assume that an optimal pulse length minimizes additional decoherence \cite{nphys4052} or that the time-dependent
coherence of the quantum emitter is small in comparison to our investigated delay times $\tau$ \cite{thoma2016exploring}. 
The coupling to the reservoir in Hamiltonian~\eqref{eq:Hfb} includes a sinusoidal dependence on the distance $\sin(\omega L/c_0)$. 
This is a non-Markovian feature induced by the reflecting mirror. 
Thus, for a simulation, a memory kernel of the non-Markovian reservoir is needed. 
To efficiently deal with the large Hilbert space, we model it within the quantum stochastic Schr\"odinger formalism, following \cite{PhysRevLett.116.093601}.
The main idea is to discretize the time evolution into equidistant time steps $t_k=k\Delta t$ and $t_{k+1}-t_k=\Delta t$.
In order to define the time discrete time-evolution operator from $t_k$ to $t_{k+1}$, we transform the feedback Hamiltonian by introducing a rotating frame with $\omega_L$ (assuming resonant excitation $\omega_L=\omega_{0}$) and defining the time-dependent bath operators 
\begin{equation}
\bv(t)=\frac{1}{\sqrt{2\pi}} \int^\infty_{-\infty} d\omega \bv(\omega) e^{-i(\omega-\omega_L)t}.
\end{equation}
The  Hamiltonian is then written as
\begin{align}
&H_{\mathrm{s,rf}}(t)=
\hbar \Omega(t) \left(  \sigma_+ + \sigma_- \right),\label{eq.:Htls,rf} \\
&H_{\mathrm {fb,rf}}(t)=-i\hbar \left( \sqrt{\frac{\Gamma}{2}} \bv(t-\tau) e^{-i\phi} +\sqrt{\frac{\Gamma}{2}} \bv(t) \right) \sigma_{+}+h.a., \notag
\end{align}
and the corresponding time-evolution operator from time $t_k$ to $t_{k+1}$
reads
\begin{equation}
U(t_{k+1},t_{k})=\hat{T}\left[ \exp\left( -\frac{i}{\hbar} \int_{t_k}^{t_{k+1}} dt' H(t') \right) \right].
\label{eq:U}
\end{equation}
In defining the photon-bin operators $\dBv(t_k)=\int_{t_k}^{t_{k+1}} dt b(t)$, which only act on the time interval $t_{k+1}-t_k$, the time-ordering operator $\hat T$ becomes redundant.
These photon-bin operators obey the commutation relations
\begin{equation}
\left[ \dBv(t_j),\dBd(t_k) \right]=\Delta t\delta_{j,k},
\label{eq:Bcomm}
\end{equation}
and we introduce the basis states 
\begin{equation}
|i_p\rangle = \frac{\left(\dBd(t_p) \right)^{i_p} }{\sqrt{i_p! \Delta t^{i_p}}}|\mathrm{vac}\rangle.
\label{eq:pbines}
\end{equation}
in the same manner as in Ref. \cite{PhysRevLett.116.093601}. %
The Schr\"odinger wave function reads in the new 
basis
\begin{align}
|\Psi\rangle= 
\sum_{\lbrace i\rbrace}
\psi_{...,i_k,i_S,...,i_{k-l},...}
\left|...,i_k,i_S,i_{k-1},...,i_{k-l},... \right\rangle \label{eq:psi}
\end{align}
with coefficient $\Psi_{...,i_k,i_S,...,i_{k-l},...}$.
However, for a time discretization of $\tau/\Delta t=100$, where $\Delta t$ is the numerical timestep, and a maximal photon number $n=4$ of the reservoir, this would correspond to a Hilbert space of approximate $3 \times 10^{60}$ states for one $\tau$-interval. 
To efficiently treat the time-evolution, we decompose $|\Psi(t_k+1)\rangle$ with a series of singular-value decompositions such that it can be written as a matrix product state. 
The singular values express the entanglement between system and reservoir. 
If singular values are sufficiently small, the state is truncated by neglecting these singular values and thus the matrix dimension is reduced \cite{Schollwoeck201196}. 
After decomposing $|\Psi(t_k+1)\rangle$, the coefficient reads
\begin{equation}
\psi_{\lbrace i \rbrace}=A^{[k]}_{i_k,\alpha_k} A^{[\mathrm{S}]}_{\alpha_k,i_S,\beta_{\mathrm{S}}} A^{[k-1]}_{\beta_{\mathrm{S}},i_{k-1},\beta_{k-1}}  \dots A^{[-l]}_{\beta_{-l},i_{k-l}}...~,\label{eq:mps}
\end{equation}
where $k$ is the future time-bin (with $i_k$ as physical index), $S$ is the tensor of the system (with $i_S$ as physical index) and $k-l$ is the feedback time-bin (with $i_{k-l}$ as physical index). 
Thus, the tensors $A$ represent either photon bins or the system. All indices $\alpha_i$ and $\beta_i$ correspond to links between the tensors.
By writing the state of the system and the reservoir in such a way, one can cut the zero value Schmidt coefficients and thus efficiently deal with a large Hilbert space. 
Initially, the state $|\Psi(0)\rangle$ represents the system in the ground state and the reservoir in a vacuum state.
Due to the pulsed excitation, it is feasible to expand the time-evolution operator to a higher order in $\Delta t$ to deal with the two different time scales of the pulsed excitation scheme.
To write the Hamiltonian in Eq. \ref{eq.:Htls,rf} in matrix form, we use the basis $|i_S,i_{n},i_{\tau}\rangle$,
where $i_S$ is the level of the TLS, $i_n$ is the occupation of the
photon bin at the current time step $t_k$ and $i_{\tau}$ is the occupation of the photon bin at time step $t_{k-l}=t_k-\tau$.
With this, we get the system matrix:
\begin{align}\notag
& 
\mathbf{M}_{\mathrm{TLS,env}}(t_n)
=
\int_{t_k}^{t_{k+1}}
\langle j_S,j_{n},j_{\tau}| H_{\mathrm{TLS,rf}}(t) |i_S,i_{n},i_{\tau} \rangle
dt\\
&=
\hbar\Delta t \big[\Omega(t_n) \left(\delta_{j_S,1} \delta_{i_S,0}+\delta_{j_S,0} \delta_{i_S,1}\right)\big] \delta_{j_n,i_n} \delta_{j_{\tau},i_{\tau}}.
\end{align}
We assume the envelope function $\Omega(t)$ to be slowly varying in the time step $\Delta t$. 
Furthermore, we use that the
system operators are not explicitly time-dependent.
The feedback reservoir matrix
is obtained via
\begin{align}
&\mathbf{M}_{\mathrm{fb}}=
\frac{i}{\hbar \sqrt{\Delta t}} \int_{t_k}^{t_{k+1}}
\langle j_S,j_{n},j_{\tau}|
H_{\mathrm{fb,rf}}  |i_S,i_{n},i_{\tau} \rangle
dt \\ \notag
=&\left( \sqrt{\frac{\Gamma}{2}} \sqrt{i_{\tau}} \delta_{j_{\tau}+1,i_{\tau}} e^{-i\phi} +\sqrt{\frac{\Gamma}{2}} \sqrt{i_{n}} \delta_{j_{n}+1,i_{n}} \right) \delta_{j_S,1}\delta_{i_S,0}  \notag\\ \notag
&-\left( \sqrt{\frac{\Gamma}{2}} \sqrt{j_{\tau}} \delta_{j_{\tau},i_{\tau}+1} e^{-i\phi} +\sqrt{\frac{\Gamma}{2}} \sqrt{j_{n}} \delta_{j_{n},i_{n}+1} \right) \delta_{j_S,0}\delta_{i_S,1}.
\end{align}
We extract the time dependency of the pulsed excitation in order to deal with time independent matrices of the system, defining the matrix $\mathbf{M}_{\mathrm{TLS}}=\mathbf{M}_{\mathrm{TLS,env}}(t_n)/\Omega(t_n)$. This has computational reasons as only the enveloping function $\Omega(t)$ changes with each time step.
When evaluating the evolution matrix in higher order, all terms of $\Delta t$ up to the desired order have to be
taken into account in the expansion
\begin{align}
\mathbf{U}&=\exp\left( \Omega(t_n) \mathbf{M}_{\mathrm{TLS}}+\mathbf{M}_{\mathrm{fb}} \right)
\nonumber\\
&= \sum_{p=0}^{\infty} \frac{1}{p!}\left( \Omega(t_n) \mathbf{M}_{\mathrm{TLS}}+\mathbf{M}_{\mathrm{fb}} \right)^p.
\label{eq:Uexp}
\end{align}
For the first order evaluation in $\Delta t$, as used in \cite{PhysRevLett.116.093601}, terms up to the order $p=2$ in the expansion of $\mathbf{U}$ contribute, as $\mathbf{M}_{\mathrm{fb}}\propto \sqrt{\Delta t}$. Thus, for second order expansion in $\Delta t$ terms up to $p=4$ in Eq. \eqref{eq:Uexp} have to be considered. We use the expansion to second order which reads explicitly
\begin{align}
\mathbf{U}&\approx \mathbf{U}_0 +\Omega(t) \mathbf{U}_1 +\Omega(t)^2 \mathbf{U}_2\notag\\
&=\mathbb{1}+\mathbf{M}_{\mathrm{fb}}+\frac{1}{2} \mathbf{M}_{\mathrm{fb}}^2+\frac{1}{6}\mathbf{M}_{\mathrm{fb}}^3+ \frac{1}{24} \mathbf{M}_{\mathrm{fb}}^4\notag\\
&+\Omega(t_n)\big[\mathbf{M}_{\mathrm{TLS}}+\frac{1}{2}\left(\mathbf{M}_{\mathrm{TLS}} \mathbf{M}_{\mathrm{fb}}+\mathbf{M}_{\mathrm{fb}} \mathbf{M}_{\mathrm{TLS}} \right)\notag\\
&+\frac{1}{6} \left( \mathbf{M}_{\mathrm{TLS}}\mathbf{M}_{\mathrm{fb}}^2+ \mathbf{M}_{\mathrm{fb}} \mathbf{M}_{\mathrm{TLS}} \mathbf{M}_{\mathrm{fb}} + \mathbf{M}_{\mathrm{fb}}^2 \mathbf{M}_{\mathrm{TLS}} \right)\big]\notag\\
&+\Omega(t_n)^2\frac{1}{2}\mathbf{M}_{\mathrm{TLS}}^2.
\end{align}
The second line is the time-independent part of the evolution matrix $\mathbf{U}_0$, in the third and fourth line
the time-dependence enters linearly and gives the linear part $\Omega(t_n) \mathbf{U}_1$. The last line is quadratic in the
pump and gives the part $\Omega(t_n)^2 \mathbf{U}_2$.
With this, the time-evolution matrices of each order can be computed
from the matrices $\mathbf{M}_{\mathrm{fb}}$ and $\mathbf{M}_{\mathrm{TLS}}$ by simple matrix multiplications.
The enveloping function $\Omega(t)$ only needs to be evaluated once each time step.
The time evolution of the system is evaluated by the sum
\begin{equation}
|\Psi(t_{k+1})\rangle = \left[ \mathbf{U}_0 +\Omega(t) \mathbf{U}_1 +\Omega(t)^2 \mathbf{U}_2 \right] |\Psi(t_k)\rangle.
\end{equation}
This can be simplified by saving the matrices $\mathbf{U}_i$ as sparse matrices so that the matrix multiplications are
only marginally slower than for the time-independent evolution.
The greatest advantage in using a higher order in $U(t_{k+1},t_k)$ is the higher possible step size $\Delta t=t_{k+1}-t_k$ with the same accuracy of the result.
Thus in total, less steps need to be performed.
In addition, a single step needs fewer singular value decompositions as $l=\tau/\Delta t$ becomes smaller and results in a high speedup of the computation.
A disadvantage of the higher order in $\mathbf{U}$ is that multi-photon processes become possible in a single time step.
Thus, additional photon states in the time-bins have to be taken into account.
However, this additional complexity is outweighed by far by the speedup due to the reduction in singular value decompositions.
%

\section{Selective-control of photon probabilities}
%
%
\begin{figure}[t!]
\centering
\includegraphics[width=\columnwidth]{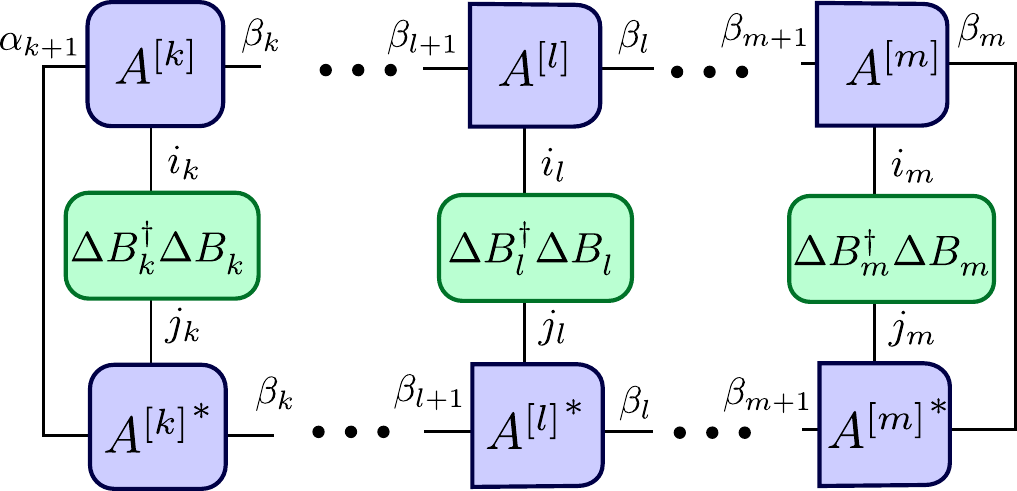}
\caption{\label{fig:MPSg3s} Computation of one integrand of $C_3$ from the matrix product state in diagrammatic form, round edges represent orthogonality. For each time combination, the intensity operator is applied at the corresponding time bins $A^{[...]}$. The cost of this operation grows linearly with the difference $|k-m|$.}
\end{figure}
Over a wide range of the pulse area $A$ of the externally applied pulse, single photon emission is the dominant process.  
However, at $A=2n\pi$, where the excitation pulse induces full Rabi-oscillations of the TLS, the two-photon probability $p(2)$ is higher than $p(1)$ \cite{nphys4052, fischer2017pulsed}. 
During the excitation pulse, the TLS might decay and emit a photon.
The remaining pulse re-excites the TLS and a second photon is emitted on a long timescale $1/\Gamma$. 
Our idea is to add an additional control parameter to steer the photon emission in this scenario and enhance just a single photon-probability, here $p(2)$.
Photon probabilities $p(n)$ are accessible via the time-integrated correlation functions:
\begin{align}
\hat I^m = 
\left( 
\sum_{j=-\infty}^{\infty} \dBd(t_j) \dBv(t_j)
\right)^m.
\end{align}
To calculate the photon probabilities from the unnormalized time integrated correlation functions \cite{1367-2630-16-5-055018}, we use the Fock state 
expansion of the photon density matrix
\begin{equation}
C_m=\langle: \hat{I}^m :\rangle = \sum_{n=0}^{\infty} \frac{n!}{(n-m)!} p(n),
\label{eq:gmbath}
\end{equation}
where $:$ indicates the normal ordering of the operators, e.g.:
\begin{align}
C_2&=\sum_{k=-q}^{N}\sum_{l=-q}^{N} \langle \dBd(t_k)\dBd(t_l)\dBv(t_l)\dBv(t_k) \rangle
\end{align}
For the numerical evaluation, we note, that there will be no light emitted into the environment before time $t=-\tau=-q\Delta t$, as we assume an initial vacuum state and after a large enough time $t_{\mathrm{end}}=N \Delta t$, all excitation from the TLS will be emitted into the bath, so that afterwards no photons will be observed.
Assuming that $p(4)$ is negligible, we yield a closed
set of equations via
\begin{subequations}
\begin{align}
p(1)&=C_1-C_2+\frac{C_3}{2},\\
p(2)&=\frac{C_2-C_3}{2},\\
p(3)&=C_3/6.
\label{eq:p3}
\end{align}
\label{eq:p(n)}
\end{subequations} 
We stay in pump regimes in which $p(3)$ is small compared to $p(1)$ and $p(2)$. 
\begin{figure}[t!]
\includegraphics[width=0.95\columnwidth]{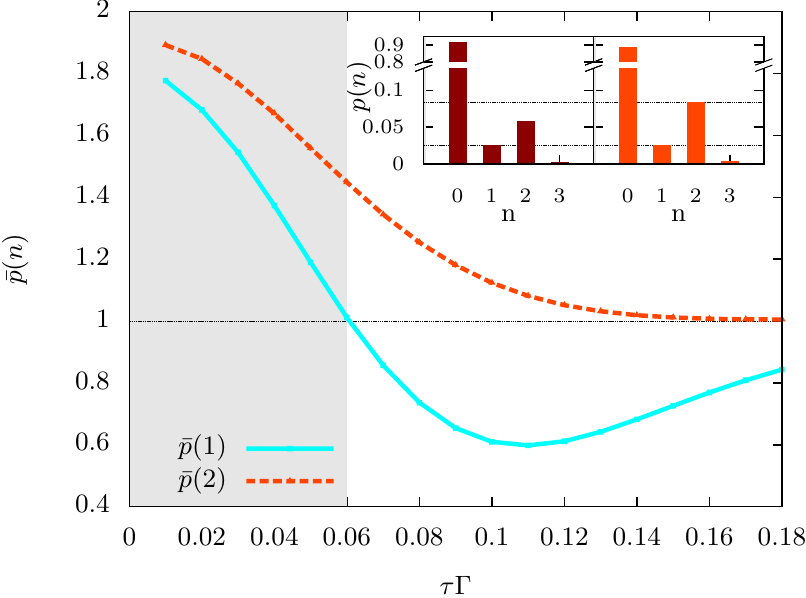}
\includegraphics[width=0.95\columnwidth]{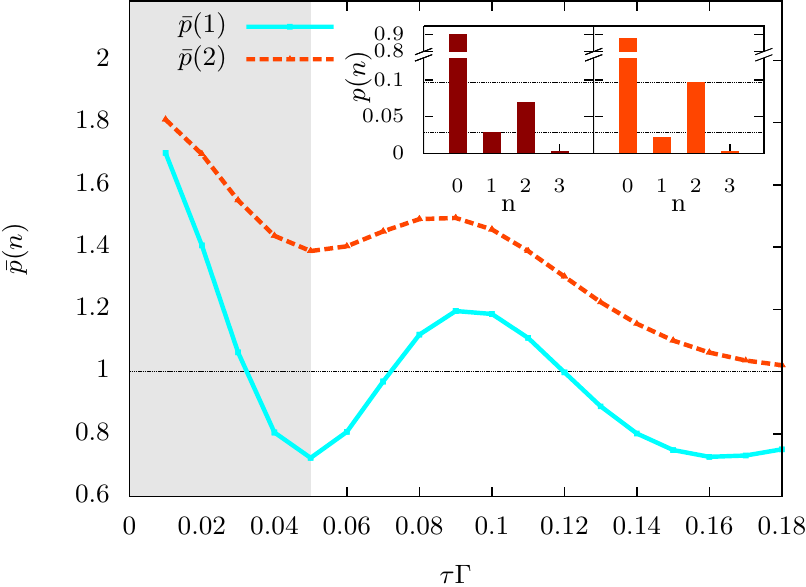}
\caption{Normalized probabilities $\bar p(n)$ ($\bar p(n)=1$ is the 
no-feedback case) for destructive feedback vs. delay time $\tau$  for $A=2\pi$ and $\nu=\frac{1}{10\Gamma}\frac{1}{\sqrt{2\text{ln}(2)}}$ (a).  Two-photon emission $\bar p(2)\geq 1$ is enhanced with feedback. Inset: $p(n)$ for no feedback (red, left) and with time-delay $\tau\Gamma=0.06$ (orange, right). $p(2)$ is increased by appr. $50\%$. For $A=4\pi$ (b), the delay $\tau$ gives more access to control the photon statistics, due to the additional Rabi-oscillation.}
\label{fig:tauplot}
\end{figure}
This allows us to assume any correlations higher than third-order to be negligible and justifies the cut-off in the expansion.
Note, the correlation functions are non-local expectation values in time and are computed from the matrix product state after the time integration. 
Thus, for the computation of the correlation functions, we need a memory kernel for all integrated time steps. 
The computation algorithm for a single integrand of $C_3$ is depicted in Fig. \ref{fig:MPSg3s} in diagrammatic form to give an example. 
The $A^{[...]}$ tensors are time-bins of the reservoir in canonical form \cite{Schollwoeck201196} at the corresponding timestep. 
According to the commutation relation in Eq. \ref{eq:Bcomm}, the correlations are invariant under the
reordering of the bath operators at different times. We can use this symmetry to reduce the cost of the numerical evaluation.  
We note that the higher order in $U(\Delta t)$ was obligatory for a numerical accessible computation of the third order correlation function.
Having the photon-probabilities at hand, we can discuss the main result of this investigation with Fig.~\ref{fig:tauplot}.
The photon-probabilities for one- $\bar p(1)$ and two-photon events  $\bar p(2)$ for a control phase of ($\phi=0$) is plotted for increasing delay time $\tau$ and normalized to the probabilities of the case without feedback: $ \bar p(n)=p(n)|_\text{feedback}/p(n)|_\text{no feedback} $.
Therefore, a value of $\bar p(n)=1 $ refers to the case in which feedback does not change the photon-probability $ p(n) $.
Remarkably, one and two-photon events depend differently on the mirror distance.
This allows to enhance two-photon events without changing the 
probability of one-photon events, cf. Fig.~\ref{fig:tauplot} at $\tau\Gamma=0.06$.
This observation motivates our claim that quantum Pyragas control gives access to manipulate individual photon probabilities 
$p(n)$, as $ p(3) $ is also not changed within numerical accuracy. 
To clarify this finding, we plot the photon-probability distribution for this case, cf. Fig.~\ref{fig:tauplot} ((a), inset),
for the case without feedback ((a), inset, left) and with feedback ((a), inset, right). 
Clearly, we address the two-photon probability without changing the one-photon probability.
This is qualitatively not expected in typical coherent quantum control setups and not within reach of Markovian quantum control, where a Lindblad dissipator governs the dynamics.
Beyond this remarkable qualitative results, quantitatively the photon-probabilities for two-photon events is enhanced by $50\%$. 
A further possibility for more control over the photon statistics is to increase the amplitude of the driving laser to a pulse area of $A=4\pi$ which we show in Fig. \ref{fig:tauplot} (b). In general, the total photon output is increased for both cases, with feedback (Fig. \ref{fig:tauplot} (b), inset, left) and without feedback (b), inset, right). As only the amplitude increases, the TLS undergoes an additional Rabi oscillation on the same time scale. Thus, the same time-delay allows for more control, e.g. at $\tau\Gamma=0.05$ we increase $p(2)$ by appr. $40\%$ and simultaneously decrease $p(1)$ by  appr. $30\%$. 
Altogether, this demonstrates that a single TLS can be used to efficiently generate a two-photon state with a high degree of control \cite{PhysRevA.96.043857, gonzalez2013two}.
Furthermore, it shows that non-Markovian quantum Pyragas control expands the possibilities to shape, tailor and manipulate individual photon-probabilities.
A decisive difference between classical and quantum Pyragas control is the phase $ \phi $, cf. Eq.~\eqref{eq:qPyragas}.
In principle, the feedback phase $ \phi=\omega_0\tau $ effectively triggers the spontaneous emission after delay $\tau$ and enhances or suppresses individual emission events. %
\begin{figure}[t!]
\includegraphics[width=\columnwidth]{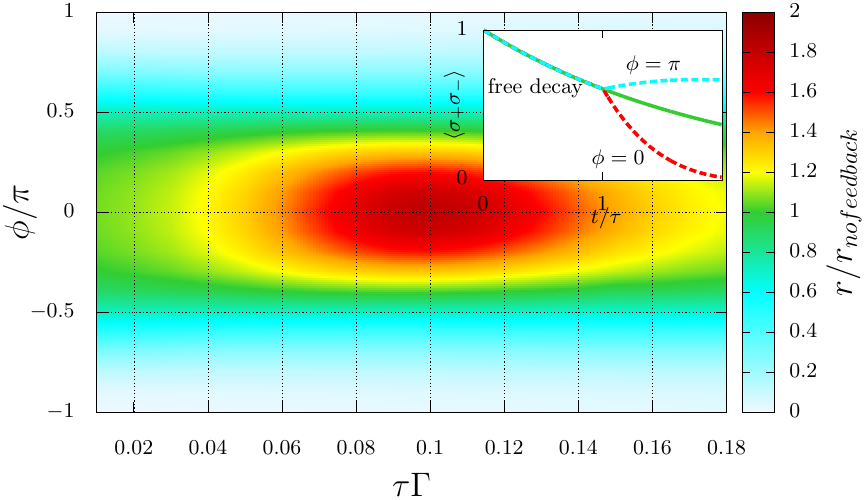}
\centering
\caption{Feedback-phase dependency of the normalized ratio $r=p(2)/p(1)$ at $A=2\pi$. $r/r_{nofeedback}=1$ indicates the case without feedback (green). $p(1)$ dominates for $\phi=\pi$ (cyan) and $p(2)$ for $\phi=0$ (red/yellow). Inset: Sketch for the TLS decay, for no- (green), constructive- (cyan) and destructive feedback (red).}
\label{fig:phivstau}
\end{figure}
If the delay is in the order of the pulse width $\nu$, 
the phase is a control parameter and different results are achieved by tuning it. 
In Fig. \ref{fig:phivstau}, we discuss the impact of the phase $\phi$ by plotting the normalized ratio $r=p(2)/p(1)$ for different delay times $\tau$ and phases $\phi$. 
If $r/r_{nofeedback}=1$ (Fig. \ref{fig:phivstau}, green), the case without feedback is reproduced. 
We observe that two-photon emission is dominant for  $\phi \in [-\pi/2,\pi/2]$ which is the destructive case where spontaneous emission is increased (see inset Fig. \ref{fig:phivstau}, red).
The influence of this phase is most easily seen in the case of
spontaneous emission with a driving field.
The analytical solution from $ [\tau,2\tau] $ reads:
\begin{align}
&\left\langle \sigma_+\sigma_-(t) \right\rangle =
e^{-2\Gamma t} \\ \notag
&
+
e^{-\Gamma (2t-\tau)}
(\Gamma t-\Gamma \tau)
\left[
2\cos(\omega_0\tau)
+
(\Gamma t-\Gamma \tau)
e^{\Gamma\tau}
\right].
\end{align}
For short delays and $ \Gamma t \ll 1 $,
the phase has a strong impact, cf.~inset Fig.~\ref{fig:phivstau}. 
For $\phi \in [\pi/2,3\pi /2]$ the feedback is constructive resulting in a suppression of spontaneous emission (cyan).
For the driven case, we note that then $p(2)$ is suppressed and $p(1)$ dominates. 
The phase $\phi$ represents fast oscillations and is more sensitive to the distance in comparison to $\tau$.
For a typical quantum dot with band gap of $1$eV, destructive interference is robust for $\Delta L \approx 0.3 \mu$m. 
For an exemplary superconducting circuit of $\omega_{0}/2\pi=6$GHz \cite{wallraff2004strong}, two photon enhancement is robust for $\Delta L \approx 1.3$cm. 
Instead of changing the distance, we propose also to change the TLS transition frequency to tune in and out of destructive interference as it is accessible in e.g. superconducting circuits \cite{wallraff2004strong,devoret2013superconducting}. 
\section{Conclusion} 
Our findings demonstrate the wide range of Pyragas control deep into the quantum regime where quantum interferences between the photon-field and the two-level system results in a higher probability of two-photon emission compared to the case without feedback while at the same time the one-photon probability is not changed. 
By using time delay $\tau$, which is tunable by the feedback geometry, as an additional control parameter, we propose a controllable setup for manipulating and tailoring feasible parts of the photon statistics which opens up new possibilities for quantum- optical spectroscopy \cite{kira2011quantum}.
For short delay times, single- and two-photon emission increase simultaneously due to a globally, on-the-fly increased decay rate. 
For a delay in the order of the pulse width, single- and two-photon emission respond differently to the feedback control. 
This allows us to achieve a two-photon enhancement up to $50\%$.  Higher pulse areas give more access to feedback-control, resulting in a more effective and pure two-photon source.\\

\section{acknowledgments}
We thank Florian Katsch for helpful discussions. The authors gratefully acknowledge the support of the Deutsche Forschungsgemeinschaft (DFG)
through the project B1 of the SFB 910 and by the School of
Nanophotonics (SFB 787).

\bibliographystyle{apsrev4-1}
\end{document}